\documentclass[11pt]{article}
\usepackage{amsmath}
\usepackage{amssymb}
\usepackage{amsthm}
\usepackage{abstract}
\usepackage{algorithm}
\usepackage{algorithmic}
\usepackage{dsfont}
\usepackage{authblk}
\usepackage{enumerate}
\usepackage{graphicx,color}
\usepackage{booktabs}
\usepackage{threeparttable}

\newtheorem{theorem}{Theorem}
\newtheorem{lemma}{Lemma}

\newtheorem{cor}[theorem]{Corollary}
\newtheorem{definition}{Definition}
\newtheorem{remark}{Remark}

\title{Combinatorial Algorithms for Partial Inverse Min-Max Spanning Tree
Problem under the Weighted Sum Hamming Distance}

\author[a]{Qingzhen Dong}
\author[a]{Xianyue Li \thanks{Corresponding author, E-mail: lixianyue@lzu.edu.cn}}
\author[a]{Xiaowen Kong}
\affil[a]{School of Mathematics and Statistics, Lanzhou University, \authorcr
	Lanzhou, Gansu, 730000, P.R. China}

\date{}
\begin{document}
	\maketitle
	\begin{abstract}
		Given a connected weighted graph $G=(V, E, w)$ and a forest $F$ of $G$,
		the goal of partial inverse min-max spanning tree problem is
		to change the weight function as little as possible,
        so that there is a min-max spanning tree with respect to the new weight function containing $F$.
        In this paper, we study this problem under the weighted	sum Hamming 
        distance. we take the alternative set of the maximum edge 
        weight 
        in $F$ with respect to the optimal weight function as a key factor, and 
        further 
        consider the case of this problem under the restriction of this factor. 
        On 
        this basis,
        a polynomial time algorithm with time complexity $O(|V|^4)$ 
        is proposed to solve this problem.\\
		
	\end{abstract}
	
		\section{Introduction}
	Given a combinatorial optimization problem $P$ (called underlying problem)
	and a partial solution $T'$
	(contained in some feasible solutions), the objective of the partial
	inverse combinatorial optimization problem is to minimize the change in
	parameter such that $T'$ is included in an optimal solution with respect to
	the new parameter. Researchers typically measure the magnitude of
	parameter changes using bottleneck-type distances ($l_{\infty}$-norm and
	bottleneck Hamming distance) and sum-type distances ($l_p$-norm and weighted sum Hamming distance).
    When $T'$ is a feasible solution, the problem degenerates into an inverse
	combinatorial optimization  problem.  The foundational research on
	 inverse combinatorial optimization problem  first appeared in 1992
	\cite{B1992},
	the paper studied inverse shortest path  problem, it
	is widely used in traffic planning and seismic tomography. Building upon
	this pioneering work, the academic community has conducted extensive
	research in this field. Primarily involving inverse maximum flow  problem
	\cite{L2006flow},
    inverse  maximum perfect matching  problem
	\cite{IMATCHING}, inverse  minimum cut problem  \cite{L2007cut}, and inverse spanning tree
	problem \cite{ IMST2, IMST4}.
    Readers can refer to the surveys \cite{D2014, H2004} and the monograph \cite{book}
    for a comprehensive understanding of inverse combinatorial optimization problems.

	Even though the partial inverse combinatorial optimization problem is
	somewhat more challenging than the inverse combinatorial optimization
	problem, researchers have
	actively pursued it and achieved significant results. Related studies have
	primarily focused on partial inverse linear programming problem
	\cite{PIliner},
	partial inverse assignment problem \cite{PIassignment}, partial inverse
	matroid problem \cite{PImatroid} and partial inverse minimum cut problem
	\cite{GPIcut}.
	
	Notably, spanning tree as a classic combinatorial optimization
	problem, researchers are particularly active in studying the partial
	inverse spanning tree problem.
	Cai et al. \cite{CPIMST} investigated the partial inverse minimum spanning
	tree problem (abbreviated as PInvMST) where weight can only be decreased
	under monotone norms. By exploring
	the construction method of optimal solution, they proved this problem is
	polynomially solvable.
	In stark contrast, Li et al. \cite{LPIMST1} studied PInvMST where weight
	can only
	be increased under the $l_p$-norm. Their work indicated that the
	problem is polynomial time solvable when the partial solution contains only
	one edge; however, it becomes APX-hard when the partial solution contains
	two or more edges. In fact, this conclusion also holds for PInvMST under the
	weighted sum Hamming distance \cite{LPIMST2}. For the situations of APX-hard,
    Li et al. \cite{LPIMST3} explored them from the perspective of approximation algorithm.
    Based on the concept of minimum fitted cut and the exact algorithms on one edge case,
    they designed approximation algorithms on PInvMST under the $l_p$-norm and the weighted sum Hamming distance.
    Specifically, they showed the approximation ratios are $\sqrt[p]{K}$ and $K$, respectively,
    where $K$ denotes the number of edges in the partial solution.
    This represents a significant advancement in the
	field of approximation algorithm for the inverse and partial inverse
	combinatorial optimization problems.
    For the bottleneck Hamming distance \cite{LPIMST2} and the $l_{\infty}$-norm \cite{LPIMST4},
    scholars have proven that PInvMST is
	polynomial solvable. For a more detailed summary,  readers can refer
	to Chapter 12 of the  monograph \cite{book}.
	The aforementioned literatures provide a systematic and comprehensive
	resolution to PInvMST.
	
	The underlying problem we study in this paper is the min-max spanning tree problem (abbreviated as MMST).
    Given a connected graph $G=(V,E)$ and an edge weight function $w$,
    a spanning tree is an MMST if its weight is the smallest among all spanning trees,
    where the weight of a spanning tree $T$ is the maximum weight of all edges in $T$,
    i.e., $w(T)	=\max_{e \in T}w(e)$.
    MMST is polynomial solvable with time complexity  $O(|E|)$ \cite{CMMST}.
    Existing research has conducted	in-depth exploration on MMST,
    encompassing both inverse MMST (abbreviated as InvMMST) and partial inverse MMST (abbreviated as PInvMMST).
    Research related to InvMMST has been discussed in \cite{L2009, L2008, Y2007},
    proving that InvMMST is	polynomial solvable under the classical norms.
    Regarding PInvMMST,	Tayyebi and Sepasian \cite{TPIMMST} first investigated 
    PInvMMST under the $l_1$-norm, and proposed two algorithms with the same 
    time complexity  $O(|E|^2|V|^2\log|V|)$ to solve this problem.
    Dong et al. \cite{DPIMMST1} proved that PInvMMST
    under the bottleneck Hamming distance is polynomial solvable with time complexity  $O(|E|\log|E|)$.
    Research on PInvMMST under the  $l_{\infty}$-norm has also advanced \cite{DPIMMST2}.
	By studying the	properties of a special class of optimal solution and the optimal value,
    and combining the decision version of this with binary search,
    this paper proposed an exact algorithm with time complexity $O(|V||E|\log|V||E|)$ to solve it.
    Thus, the main challenge in completely solving PInvMMST
    under the classical norms lies in the weighted sum Hamming distance.
    In this paper, we will explore PInvMMST under the weighted sum Hamming distance,
    bringing the research on PInvMMST to a perfect	conclusion.
    Table \ref{TABLE} fully shows the computational	complexity of PInvMMST under the classical norms.

    \begin{table}[htbp]\label{TABLE}
		\centering
		\begin{threeparttable}
			\caption{The result on the computational complexities for PInvMMST,
			where BH stands for the bottleneck Hamming distance and WSH stands for
			the weighted sum Hamming distance}
			\begin{tabular}{lcccc}
				\hline
				 & $l_1$-norm & BH & $l_\infty$-norm & WSH \\
				\hline
				PInvMMST & $O(|E|^2|V|^2\log|V|)$ [22] & $O(|E|\log|E|)$ [7] &
				$O(|V||E|\log|V||E|)$ [6] & $^\sharp$ $O(|V|^4)$
			\\
				\hline
			\end{tabular}
			\begin{tablenotes}
				\item[$^\sharp$ ] This result is obtained in this paper
			\end{tablenotes}
		\end{threeparttable}
	\end{table}

	The rest of this paper will be studied according to the following structure.
    In Section \ref{Sec:PR},
    we provide a formal definition of PInvMMST under the weighted sum Hamming distance,
    present a condition for determining the existence of feasible solutions to the problem,
	and derive a property of the optimal solution.
    Subsequently, in Section \ref{ALT}, we focus on a key factor 
    $w^*(F)$, 
    which is the maximum edge weight in $F$ with respect to the optimal 
    weight function $w^*$. We identify a set of values of $w^*(F)$ that play a 
    crucial role in determining the optimal solution, and define this set as 
    the alternative set of $w^*(F$).
    This theoretical result serves as the foundation for the algorithm design in Section \ref{RES}.
    Section \ref{RES} firstly discusses the case of this problem with the 
    restriction of $w^*(F)$,  
    and provides a polynomial time algorithm to solve it;
    and then based on this algorithm,
    a polynomial time algorithm for PInvMMST under the weighted sum Hamming 
    distance has been presented.
    Section \ref{conclusion} summarizes the research findings and provides conclusions.
	
	\section{Preliminary}\label{Sec:PR}
	
	In this section, we will provide the definition of \textbf{partial inverse
	min-max spanning tree problem under the weighted sum Hamming distance
	(PInvMMST$_{WSH}$)}, present a condition for
    determining the existence of feasible solutions to the problem, and derive
    a property of the optimal solution.
	
	\begin{definition}\label{def1}
		Given a connected graph $G=(V,E)$, a forest $F$ of $G$, an edge weight function $w$,
        a norm function $c$: $E \rightarrow \mathbb{R^+} $,
        and non-negative lower and upper bound functions $l$ and $u$,
        the goal of	\textbf{PInvMMST$_{WSH}$} is to find a new weight function $w^*$ satisfying:
		\begin{enumerate}[(1)]
			\item there exists a min-max spanning tree $T^*$ of $G$ with
			respect to $w^*$ such that $F\subseteq T^*$;
			\item $-l(e)\leq w^*(e)-w(e)\leq u(e)$, $\forall e \in E$;
			\item  $\|w^*-w\|_{WSH}= \sum_{e \in E} c(e)H(w^*(e), w(e))$ is minimum;
		\end{enumerate}
	   where $H(w^*(e), w(e))=0$ if $w^*(e)=w(e)$, $H(w^*(e), w(e))=1$ if $w^*(e)\neq w(e)$.
	\end{definition}
    \noindent
    Furthermore, we call condition (2) the capacity constraint.
    If a weight function $w^\prime$ satisfies conditions (1) and (2),
    then we call $w^\prime $ a \emph{feasible solution} of PInvMMST$_{WSH}$.

    As described above, mathematical model of PInvMMST$_{WSH}$ can be constructed as follows,
    \[
        \begin{aligned}
 	  & \min \|w^* - w\|_{WSH} \\
 	  (PInvMMST_{WSH}) \quad \text{ s.t. } & \max_{e \in T^*} w^*(e) \leq \max_{e \in
 	  T} w^*(e), \quad F \subseteq T^* \in \mathcal{T} \text{ and }
 	  \forall T \in \mathcal{T}, \\
 	  & w - l \leq w^* \leq w + u.
        \end{aligned}
    \]
    where $\mathcal{T}$ is the set of all spanning trees of $G$,

    The following theorem provides a necessary and sufficient condition for
    determining the existence of a min-max spanning tree of $G$ containing $F$.
    For the convenience of the following description,
    for any weight function $w$ and any real number $p$,
    we define $E_{w,p}^{\geq} =\{{e\in E} | {w(e)\geq p}\}$, that is,
    $E_{w, p}^{\geq}$ is the set of edges whose weight is greater than or equal to $p$.
    For any edge subset $E' \subseteq E$, we define $w(E')= \max\limits_{e\in E'}w(e)$.

	\begin{theorem}\label{the1}\cite{DPIMMST1}
		Let $G=(V, E)$ be a connected graph, $F$ be a forest and $w$ be an edge
		weight function. There exists a min-max spanning tree of $G$ with
		respect to $w$ containing $F$ if and only if $G-E_{w, w(F)}^{\geq} $ is
		disconnected.
	\end{theorem}

    By Theorem \ref{the1}, the following corollary is naturally obtained.

	\begin{cor}\label{coro2}
		Given a connected graph $G=(V, E)$, a forest $F$ of $G$,
        and two edge weight functions $w$ and $w'$.
        If there exists a min-max spanning tree	with respect to $w$ containing $F$
        and $E_{w,w(F)}^{\geq} \subseteq E_{w',w'(F)}^{\geq}$,
        then there also exists a min-max spanning tree with	respect to $w'$ containing $F$.
	\end{cor}

    \begin{proof}
        Since there exists a min-max spanning tree with respect to $w$ containing $F$,
        by Theorem \ref{the1}, $G-E_{w,w(F)}^{\geq}$ is  disconnected.
        Furthermore, $G-E_{w',w'(F)}^{\geq}$ is also disconnected,
        because of $E_{w,w(F)}^{\geq} \subseteq E_{w',w'(F)}^{\geq}$.
        Again, by Theorem \ref{the1},
        there exists a min-max spanning tree with respect to $w'$ containing $F$.
    \end{proof}

	The following lemma explores the properties of the optimal solution of
	PInvMMST$_{WSH}$.
	\begin{lemma}\label{lem1}
	Let $I=(G, F, w, c, l, u)$ be an instance  of PInvMMST$_{WSH}$.
	Then for any optimal solution $w^*$ of $I$ and any edge $e\notin F$,
	$w^*(e)\geq w(e)$.
	\end{lemma}
	
	\begin{proof}
	In fact, if there exists an edge $e' \notin F$ with $w^*(e')<w(e')$. We can
	construct a new weight function $w_o$ as follows,
		\begin{equation}
			w_o(e)=\left\{
			\begin{array}{lcl}
				w(e^\prime), &  & e=e^\prime;\\
				w^*(e) , &  &  e\neq e^\prime.\\
			\end{array}
			\right.\notag
		\end{equation}
		Obviously,  $w^*$ and $w_o$ differ only on the edge $e'$,
        and $H(w_o(e'), w(e'))=0<1=H(w^*(e'), w(e'))$.
	    Therefore,
		$$\|w_o-w\|_{WSH}<\|w^*-w\|_{WSH}.$$
	    This implies that if $w_o$ is a	feasible solution of $I$,
        this will contradicts with the optimality of $w^*$.
		
		Next, we shall show the feasibility of $w_o$.
	    Obviously, $w_o$ satisfies the capacity constraint.
		Since $e' \notin F$, $w_o(F)=w^*(F)$. At the same time,
		$w_o(e')=w(e')>w^*(e')$. Therefore,
		$E_{w^*,w^*(F)}^{\geq} \subseteq
		E_{w_o,w_o(F)}^{\geq}$. By Corollary \ref{coro2},
	    there exists a min-max spanning tree with respect to $w_o$
		containing $F$.
		In summary,  $w_o$  is a feasible solution of $I$.
	\end{proof}

	\section{Alternative set of $w^*(F)$}\label{ALT}

   Through Theorem \ref{the1}, we know that the maximum value of the edge 
   weight in $F$ is a key factor
    in order to meet the first condition in Definition 1,
    so we can regard this value as a baseline.
    In this section, for any instance $I=(G, F, w, c, l, u)$ of PInvMMST$_{WSH}$,
    we will explore the value range of $w^*(F)$ determined by an optimal solution $w^*$ of $I$,
    so as to lay the foundation for finding the optimal solution of $I$.
    The following theorem describes the possible values of $w^*(F)$ in $I$,
    which we also refer to as the alternative set of $w^*(F).$

	\begin{theorem}\label{the2}
		For any instance $I=(G, F, w, c, l, u)$ of PInvMMST$_{WSH}$,  there
		exists an optimal solution $w^*$ such that the value of $w^*(F)$  is
		certainly contained within the set $W=\left\lbrace w(e)|e
		\in F \right\rbrace \cup  \underline{w}$,
        where $\underline{w}=\max \left\lbrace w(e)-l(e) |e \in
		F\right\rbrace$.
	\end{theorem}

    \begin{proof}
    	Let $w^*$ be an optimal solution of $I$.
        If  $w^*(F) \in W$, then Theorem \ref{the2} already holds.
        Otherwise, let $W'=W  \cup \overline{w}$,
        where $\overline{w}=\max \left\lbrace w(e)+u(e)|e \in F\right\rbrace +1$.
        We sort the elments in $W'$ with $w_1<w_2< \cdots <w_t= \overline{w} \left( t \leq n+1\right)$.
        Therefore, $w_i < w^*(F) <w_{i+1}$ for some $1 \leq i \leq t-1$.
        Furthermore, due to the capacity constraint, $\underline{w} \leq w^*(F)<\overline{w}$.
        Now, a new weight function $w_1$ is defined as follows,
        \begin{equation}
	       w_1(e)=\left\{
	       \begin{array}{lcl}
		      w_i, &  &e \in F  \mbox{ and } w_i < w^*(e) <w_{i+1};\\
 		      w^*(e), &  &\mbox{otherwise.}\\
	       \end{array}
            \right.\notag
        \end{equation}
        It is not difficult to see that the difference between $w_1$ and $w^*$
        is only that the weights of the edges in $F$.
        Next, we will prove that $w_1$ is still a feasible solution of $I$.

        Firstly, we prove that $w_1$ satisfies the capacity constraint.
        For any edge $e \in F$ with $w_i < w^*(e) <w_{i+1}$,
        if $0< w^*(e)-w(e) \leq u(e)$, then
        $$0 \leq w_1(e)-w(e)=w_i -w(e) < w^*(e)-w(e) \leq u(e).$$
        If $-l(e) \leq  w^*(e)-w(e)< 0$, we claim that $\underline{w} \leq w_i$.
        Otherwise, let $\bar{e}=\arg \max \left\lbrace w(e)-l(e) |e \in F\right\rbrace$.
        It can be concluded that
        $$w^*(\bar{e}) \geq w(\bar{e})-l(\bar{e})=\underline{w} \geq w_{i+1}>w^*(F),$$
        a contradiction.
        Moreover, we obtain that
        $$-l(e) \leq \underline{w}-w(e) \leq w_i-w(e)= w_1(e)-w(e)<0.$$
        Hence, for any other edge $e$, it is clear that
        $$-l(e) \leq w^*(e)-w(e)=w_1(e)-w(e) \leq u(e).$$

        Then, we prove that $w_1(F)=w_i$.
        For any edge $e \in F$, if $w^*(e) \leq w_i$, then $w_1(e)=w^*(e) \leq w_i$;
        if $w_i < w^*(e) \leq w^*(F)$, then $w_1(e)=w_i$.
        Thus, $w_1(F)=w_i<w^*(F)$.

        Finally, we prove that there exists a min-max spanning tree with respect to $w_1$ containing $F$.
        For any edge $e \in F$ with $ w^*(e) = w^*(F)$, it can obtain that $w_1(e)=w_i=w_1(F)$.
        For any other edge $e$ with $w^*(e) \geq w^*(F)$, $w_1(e)=w^*(e) \geq w^*(F) >w_1(F)$.
        Therefore, $E_{w^*, w^*(F)}^{\geq} \subseteq E_{w_1, w_1(F)}^{\geq}$.
        By Corollary \ref{coro2}, there exists a min-max spanning tree with respect to $w_1$ containing $F$.

        In summary, $w_1$ is a feasible solution of $I$.
        For any edge $e \in F$ with $w_i < w^*(e)<w_{i+1}$, easy to see that $H(w^*(e), w(e))=1 \geq H(w_1(e), w(e))$. Therefore, 	
        $$\|w_1-w\|_{WSH} \leq \|w^*-w\|_{WSH}.$$
        According to the optimality of $w^*$, $w_1$ is also an optimal solution of $I$ and $w_1(F)=w_i$.
 \end{proof}

 \begin{remark}
    From above proof, we can see that $w^*(F)\ge \underline{w}$.
    Otherwise, $w^*(F)<\underline{w}=w(\bar{e})-l(\bar{e}) \leq w^*(\bar{e})$,
    which contradicts with $w^*(F)=\max\limits_{e\in F}w^*(e)$.
    Therefore, we can obtain a more precise alternative set $W_f$
    by removing all elements in $W$ that are less than $\underline{w}$,
    that is, $W_f = W \cap [\underline{w}, +\infty).$
 \end{remark}

\section{Algorithms for PInvMMST$_{WSH}$}\label{RES}
    In Section \ref{ALT}, we have explored the alternative set of $w^*(F)$.
    Then, a natural idea is to solve PInvMMST$_{WSH}$ when $w^*(F)$ is fixed.
    For clarity, we refer to this problem as restricted version of PInvMMST$_{WSH}$.
    If restricted version of PInvMMST$_{WSH}$ is solved, by Theorem \ref{the2},
    we can solve restricted version with the values in $W_f$ one by one
    to find the optimal value and optimal solution of PInvMMST$_{WSH}$.
    Therefore, at the beginning, we will explore how restricted version of PInvMMST$_{WSH}$ is solved.

\subsection{Restricted Version of PInvMMST$_{WSH}$}

    To understand and describe restricted version of PInvMMST$_{WSH}$ more clearly,
    we firstly provide its formal definition.
	
    \begin{definition}
		Given a connected graph $G=(V,E)$, a forest $F$ of $G$, an edge weight function $w$,
        a norm function $c$: $E \rightarrow \mathbb{R^+}$,
		non-negative lower and upper bound functions $l$ and $u$, and a real
		number $p^* \in W_f$,
        the goal of \textbf{restricted version of PInvMMST$_{WSH}$}
        is to find a new weight function $w^*$ satisfying:
		\begin{enumerate}[(1)]
			\item there exists a min-max spanning tree $T^*$ of $G$ with respect to $w^*$ such that $F\subseteq T^*$;
			\item $-l(e)\leq w^*(e)-w(e)\leq u(e)$, $\forall e \in E$;
			\item $w^*(F)=p^*$;
			\item  $\|w^*-w\|_{WSH}= \sum_{e \in E} c(e)H(w^*(e), w(e))$ is	minimum.
		\end{enumerate}
	\end{definition}
	
    \noindent
    It is readily apparent that, in comparison with the definition of PInvMMST$_{WSH}$,
    an additional condition (3) has been added.
    We denote this problem as PInvMMST$_{WSH}^{p^*}$.
    If a weight	function $w^\prime$ satisfies conditions (1), (2) and (3),
    then we	call $w^\prime $ a feasible solution of PInvMMST$_{WSH}^{p^*}$.

   In order to address PInvMMST$_{WSH}^{p^*}$,
   we firstly classify the edge set $E$ into three subsets as follows,
  \begin{enumerate}[(1)]
        \item $E_1=F_{p^*}^{>} \cup F_{p^*}^{=}$,
        where $F_{p^*}^{>}=\left\lbrace e \in F | w(e)>p^*\right\rbrace$ and
        $F_{p^*}^{=}=\left\lbrace e \in F| w(e)=p^*\right\rbrace$;
   	    \item  $E_2=\left\lbrace e \in E \setminus F | w(e) \ge
   	    p^*\right\rbrace$;
   	   	\item  $E_3=\left\lbrace e \in E | w(e)<p^*\right\rbrace$.
   \end{enumerate}

    \noindent
    From the classification, we can see that $E_1 \cup E_2=E_{w, p^*}^{\geq}$.
    Moreover, since $p^*\in W_f$, if $p^*=\underline{w}$, $F_{p^*}^{>}\neq \emptyset$;
    otherwise $F_{p^*}^{=}\neq \emptyset$.
    Thus, it can conclude that $E_1 \neq \emptyset$.
    Clearly, in order to satisfy condition (3) of Definition 2,
    the weights of all edges in $E_1$ should be equal to $p^*$.
    Next, based on this modification,
    we will consider how to modify the new weight
    to satisfy condition (1) of Definition 2 as small as possible.

    Define $G(p^*)=(V, E_3)$.
    If $G(p^*)$ is disconnected, by Theorem \ref{the1}, condition (1) is already satisfied.
    If $G(p^*)$ is connected,
    then it is necessary to increase the weights of some edges in $E_3$ to $p^*$,
    so that after deleting these edges in $G(p^*)$, the new graph is no longer connected.
    Based on this analysis, the current problem can be transformed into minimum cut problem.
    Thus, we will provide the definition of minimum cut problem,
    and determine the capacities of all edges in $G(p^*)$.

    \begin{definition}
        Given a graph $G=(V, E)$ with a non-negative capacity on each edge,
        the minimum cut problem on $G$ is to find a set of edges whose deletion makes $G$ disconnected,
        and the total capacity of the edges in this set is minimized.
    \end{definition}

    Then, we define the capacity function $d$ on $E(G(p^*))$ as follows,
    \begin{equation}\label{equ1}
	   d(e)=\left\{
	   \begin{array}{lcl}
		  c(e), &  & e \in E_3 \mbox{ and } w(e)+u(e)\geq p^*;\\
		  \infty , &  & e \in E_3 \mbox{ and } w(e)+u(e)< p^* .\\
	   \end{array}
	   \right.
    \end{equation}

    Now, we present the Alg. \ref{alg:p^*} for PInvMMST$_{WSH}^{p^*}$.
    To present the algorithm more clearly, we first provide some symbolic definitions.
    For each edge set $E'$ we define  $c^S(E')=\sum_{e \in E'} c(e)$ and
    use similar notation $d^S(E')$ for capacity  $d$ (where $S$ stands for ``Sum").

    \renewcommand{\thealgorithm}{1}
    \begin{algorithm}[!http]
	\caption{Algorithm for PInvMMST$_{WSH}^{p^*}$.}\label{alg:p^*}
	\begin{algorithmic}[1]
		\REQUIRE An intance $I=(G, F, w, c, l, u; p^*)$  of	PInvMMST$_{WSH}^{p^*}$.
		\ENSURE Optimal solution $w^*$ and optimal value $opt$ or ``No Feasible	Solution!".
		\STATE Divide the edge set $E$ into $E_1$, $E_2$ and $E_3$, and determine the graph $G(p^*)=(V, E_3)$;
		\IF {$G(p^*)$ is disconnected}
		\RETURN Optimal solution $w^*$ with
		\begin{equation}\label{equ2}
			w^*(e)=\left\{
			\begin{array}{lcl}
				p^*,  &  &  e \in F_{p^*}^{>};\\
				w(e), &  &  \mbox{otherwise},\\
			\end{array}
			\right.
		\end{equation}
		and optimal value $opt=c^S(F_{p^*}^{>})$;
		\ELSE
		\STATE Add  capacity to each edge of graph $G(p^*)$ by Eqn.
		(\ref{equ1});
		\STATE Find a minimum cut $C^*$ of the graph $G(V, E_3, d)$;
		\IF {the capacity of $C^*$ satisfies $d^S(C^*)=\infty$;}
		\RETURN ``No Feasible Solution!";
		\ELSE
		\RETURN Optimal solution $w^*$ with
		\begin{equation}\label{equ3}
			w^*(e)=\left\{
			\begin{array}{lcl}
				p^*,  &  &  e \in F_{p^*}^{>}\cup C^*;\\
				w(e), &  &  \mbox{otherwise},\\
			\end{array}
			\right.
		\end{equation}
		and optimal value $opt=c^S(F_{p^*}^{>})+c^S(C^*)$.
		\ENDIF
		\ENDIF
	\end{algorithmic}
    \end{algorithm}

    \begin{theorem}\label{the3}
		Algorithm \ref{alg:p^*} is a strongly polynomial algorithm for PInvMMST$_{WSH}^{p^*}$.
    \end{theorem}

    \begin{proof}
    Firstly, we show that if $G(p^*)$ is disconnected,
    the new weight $w^*$ obtained by Eqn. (\ref{equ2}) is an optimal solution
    and $c^S(F_{p^*}^{>})$ is the optimal value of $I$.
    By the definition of $w^*$ and $E_1 \neq  \emptyset$, we have $w^*(F)=p^*$.
    Furthermore, for any edge $e \in F_{p^*}^{>},$
    \begin{equation}
	   \begin{aligned}
		  0< w(e)-w^*(e) & =w(e)-p^* \\
		                 & \leq w(e)- \underline{w} \\
		                 & \leq w(e)-\left( w(e)-l(e)\right) \\
		                 & =l(e).
	   \end{aligned} \notag
    \end{equation}
    Thus, $w^*$ satisfies the capacity constraint.
    At the same time, $G(p^*)$ being disconnected indicates that
    there exists a min-max spanning tree of $G$ with respect to $w^*$ containing $F$.
    Thus, $w^*$ is a feasible solution of $I$.

    The following will prove the optimality of $w^*$.
    For any feasible solution $w'$ of $I$,
    the weights of edges in $F_{p^*}^{>}$ must be modified because of $w'(F)=p^*$.
    Therefore, $$\|w'-w\|_{WSH} \geq c^S(F_{p^*}^{>})=\|w^*-w\|_{WSH}.$$
    This implies that $w^*$ is an optimal solution and $c^S(F_{p^*}^{>})$ is the optimal value of $I$.

    Next, we will prove that if $G(p^*)$ is connected,
    and the algorithm returns ``No Feasible Solution!",
    instance $I$ doesn't have feasible solutions.
    Suppose to the contrary that $I$ has a feasible solution $w'$.
    By Theorem \ref{the1},
    $G-E_{w', p^*}^{\geq}$ is disconnected.
    On the other side,
    \begin{eqnarray*}
        &   & G(p^*)-\left[ E_{w', p^*}^{\geq} \cap E_3 \right] \\
        & = & G - (E_1 \cup E_2) - \left[ E_{w', p^*}^{\geq} \cap E_3 \right] \\
    & \subseteq & G - \left[ E_{w', p^*}^{\geq} \cap (E_1 \cup	E_2)\right] - \left[ E_{w', p^*}^{\geq} \cap E_3 \right] \\
        & = & G - E_{w', p^*}^{\geq}.
    \end{eqnarray*}
    This indicates that $E_{w', p^*}^{\geq} \cap E_3$ is a cut of $G(p^*)$.
    Since $w'$ is a feasible solution, for any edge $e\in E_{w', p^*}^{\geq} \cap E_3$,
    $w(e)+u(e) \geq w'(e) \geq p^*$.
    Hence, $d(e)< \infty$ for any edge $e\in E_{w', p^*}^{\geq} \cap E_3$. Moreover,
    $$d^s(E_{w', p^*}^{\geq} \cap E_3) < \infty=d^s(C^*).$$
    This contradicts the fact that $C^*$ is the minimum cut of $G(p^*)$.

    Then, we will prove that if $G(p^*)$ is connected,
    and the algorithm returns $w^*$ obtained by Eqn. (\ref{equ3}),
    then $w^*$ is an optimal solution and $c^S(F_{p^*}^{>})+c^S(C^*)$ is the optimal value of $I$.
    By the definition of $w^*$ and $E_1 \neq  \emptyset$, we have $w^*(F)=p^*$.
    Similar to the case where $G(p^*)$ is disconnected,
    for any edge $e \in F_{p^*}^{>}$, the capacity constraint is satisfied.
    For any edge $e \in C^*$, since $d^S(C^*)<\infty$, we have
    \begin{equation*}
		0 < w^*(e)-w(e) = p^*-w(e) \leq \left( w(e)+u(e)\right) - w(e) = u(e).
    \end{equation*}
    Thus, $w^*$ satisfies the capacity constraint.
    Furthermore, since $E_{w^*, p^*}^\geq=C^* \cup E_1 \cup E_2$,
    $G-E_{w^*, p^*}^{\geq}=G(p^*)-C^*$.
    This implies that $G-E_{w^*, p^*}^{\geq}$ is disconnected.
    In summary, $w^*$ is a feasible solution of $I$ and the objective value is $c^S(F_{p^*}^{>})+c^S(C^*).$

    Let $w'$ be any feasible solution of $I$ and $C'=E_{w', p^*}^{\geq} \cap E_3$.
    Then, the weights of edges in $F_{p^*}^{>}$ must be modified because of $w'(F)=p^*$.
    Similar to the case where the algorithm returns ``No Feasible Solution!",
    $C'$ is a cut of $G(p^*)$. Thus,
    $$ \|w^*-w\|_{WSH} = c^S(F_{p^*}^{>})+c^S(C^*) \leq c^S(F_{p^*}^{>})+c^S(C') \leq \|w'-w\|_{WSH}.$$
    Therefore, $w^*$ is an optimal solution and $c^S(F_{p^*}^{>})+c^S(C^*)$ is the optimal value of $I$.

    Finally, we analyze the time complexity of Algorithm \ref{alg:p^*}.
    It is not difficult to see that each line takes $O(|E|)$ time, except for Line 6.
    However, to find a minimum cut of a graph costs $O(|V|^3)$ time \cite{Flow1993}.
    Hence, the time complexity of Algorithm \ref{alg:p^*} is $O(|V|^3)$.
\end{proof}

\subsection{Algorithm for PInvMMST$_{WSH}$}

    Combining Theorem \ref{the2} with Alg. \ref{alg:p^*},
    we provide the Alg. \ref{alg:PU} for PInvMMST$_{WSH}$.
 	
 	\renewcommand{\thealgorithm}{2}
 	\begin{algorithm}[!http]
    \caption{Algorithm for PInvMMST$_{WSH}$.}\label{alg:PU}
    \begin{algorithmic}[1]
        \REQUIRE An instance $I=(G, F, w, c, l, u)$ of PInvMMST$_{WSH}$.
       	\ENSURE Optimal solution $w^*$ and optimal value $opt$ or ``No Feasible Solution!".
       	\STATE Determine the alternative set $W_f$ and order the values in $W_f$ as $p_1 < p_2 < ......<p_t$;
 	    \STATE Initialize $opt \gets \infty$ and $w^* \gets \text{NULL}$;
 		\FOR{$i \gets 1$ \TO $t$}
 		\STATE Execute Alg. \ref{alg:p^*} on $(G, F, w, c, l, u; p_i)$;
 			\IF{Alg. \ref{alg:p^*} returns optimal solution $w^*_i$ and optimal value $opt_i$;}
                \IF {$opt_i<opt$}
 				\STATE $opt \gets opt_i$ and $w^* \gets w^*_i$;
 				\ENDIF
            \ENDIF
 		\ENDFOR
 		\IF{$opt = \infty$}
 			\RETURN ``No Feasible Solution!'';
 		\ELSE
 			\RETURN $w^*$ and $opt$.
 		\ENDIF
 	\end{algorithmic}
 	\end{algorithm}
 		
    \begin{remark}
        The time complexity of Algorithm \ref{alg:PU} is $O(|V|^4).$
    \end{remark}
	
    \begin{remark}
        Interestingly, we found when $p^*$ is an arbitrary real number,
        solving PInvMMST$_{WSH}^{p^*}$ will become more difficulty.
        The primary is that $E_1$ may be an empty set, so it cannot guarantee that $w^*(F)=p^*$.
        Indeed, when $E_1=\emptyset$, $G(p^*)$ is connected and $C^*\cap F= \emptyset$,
        the weight function $w^*$ obtained by Eqn. (\ref{equ3}) is not a feasible solution,
        since $w^*(F)<p^*$.
        Now, it is necessary to increase the weight of some edge in $F'$ to
        $p^*$ to meet $w^*(F)=p^*$,
        where $F'=\{e\in F \:|\: w(e)+u(e)\geq p^*\}$.
        However, this does not guarantee that the modified weight is optimal.
        Therefore, a natural idea at this time is to increase the weight of edge in $F'$ to $p^*$ one by one,
        find the minimum cut after $G(p^*)$ deletes this edge,
        and choose the best one in all possible cases as the output.
        However, since our current focus is on addressing PInvMMST$_{WSH}$,
        we do not elaborate further on PInvMMST$_{WSH}^{p^*}$ when $p^*$ is an arbitrary real number.
    \end{remark}

\begin{figure}[ht]
	   \centering
	   \includegraphics[width=3.5in]{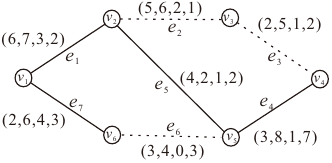}\\
	   \caption{An instance of PInvMMST$_{WSH}^{p^*}$}\label{Fig:instance}
    \end{figure}

    In the following, we give an example to explain the operation process of Alg. \ref{alg:PU}.
    Fig. \ref{Fig:instance} illustrates an instance of PInvMMST$_{WSH}$,
    in which the dash edges $e_2$, $e_3$ and $e_6$ form the partial solution $F$;
    the four numbers on each edge $e$ represent
    its weight $w(e)$, norm $c(e)$, lower bound $l(e)$ and upper bound $u(e)$, respectively.

    After running Line 1 in Alg. \ref{alg:PU}, we can obtain that $\underline{w}=3$ and $W=\{2, 3, 5\}$.
    Furthermore, $W_f=\{3, 5\}$, $p_1=3$ and $p_2=5$.
    Then, through initialization, optimal value $opt=\infty$ and optimal solution $w^*=NULL$.
    Next, Alg. \ref{alg:PU} will execute Alg. \ref{alg:p^*} twice with $p_1=3$ and $p_2=5$, respectively.

\begin{figure}[ht]
	   \centering
	   \includegraphics[width=5.5in]{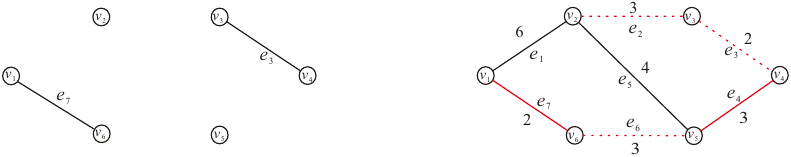}\\
	   \caption{The first call of Alg. \ref{alg:p^*} with $p_1=3$}\label{Fig:result1}
    \end{figure}

    When Alg. \ref{alg:p^*} is called for the first time, $p_1=3$.
    At this time, the implementation of Alg. \ref{alg:p^*} is divided into the following three steps.
    \begin{itemize}
        \item Step 1: Divide the edge set $E$ into 3 parts.
        In detail, $E_1=\left\lbrace e_2, e_6\right\rbrace $,
        where $F_{p_1}^{>}=\left\lbrace e_2\right\rbrace$ and $F_{p_1}^{=}=\left\lbrace e_6\right\rbrace$;
        $E_2=\left\lbrace e_1, e_4, e_5\right\rbrace $ and $E_3=\left\lbrace e_3, e_7\right\rbrace$.
        \item Step 2: Determine the graph $G(p_1)$ as the left figure of Fig. \ref{Fig:result1}.
        \item Step 3: Since $G(p_1)$ is disconnected,
        obtain optimal solution $w_1^*=(6, 3, 2, 3, 4, 3, 2)$ by Eqn.(\ref{equ2}),
        and the optimal value $opt_1=c(e_2)=6$.
        The right figure of Fig. \ref{Fig:result1} illustrates the optimal solution $w_1^*$,
        and the min-max spanning tree with respect to $w_1$ (red edges).
    \end{itemize}
    Because $opt_1<opt$, optimal solution $w^*$ is updated to
   $w^*_1$ and optimal value $opt$ is updated to 6.

\begin{figure}[ht]
	   \centering
	   \includegraphics[width=5.5in]{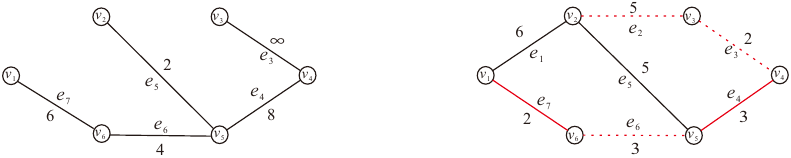}\\
	   \caption{The second call of Alg. \ref{alg:p^*} with
	   $p_2=5$}\label{Fig:result2}
    \end{figure}

    When Alg. \ref{alg:p^*} is called for the second time, $p_2=5$.
    At this time, the implementation of Alg. \ref{alg:p^*} is divided into the following four steps.
    \begin{itemize}
        \item Step 1: Divide the edge set $E$ into 3 parts.
        In detail, $E_1=\left\lbrace e_2\right\rbrace $,
        where $F_{p_2}^{>}=\emptyset$ and $F_{p_2}^{=}=\left\lbrace e_2\right\rbrace$;
        $E_2=\left\lbrace e_1\right\rbrace $ and $E_3=\left\lbrace e_3, e_4, e_5, e_6, e_7\right\rbrace$.
        \item Step 2: Determine the graph $G(p_2)$ as the left figure of Fig. \ref{Fig:result2}.
        \item Step 3: Since $G(p_2)$ is connected, determine the capacity of each edge by Eqn.(\ref{equ1})
        (See the numbers above the edges in the left figure of Fig. \ref{Fig:result2}),
        and find a minimum cut $C^*=\left\lbrace e_5\right\rbrace$ of $G(p_2)$.
        \item Step 4: Obtain optimal solution $w_2^*=(6, 5, 2, 3, 5, 3, 2)$ by Eqn.(\ref{equ3}),
        and the optimal value $opt_2=c(e_5)=2$.
        The right figure of Fig. \ref{Fig:result2} illustrates the optimal solution $w_2^*$,
        and the min-max spanning tree with respect to $w_2$ (red edges).
    \end{itemize}
    Since $opt_2<opt$, optimal solution $w^*$ is updated to $w^*_2$ and optimal value $opt$ is updated to 2.

    Finally, since $opt=2$, Alg. \ref{alg:PU} outputs $w^*=(6, 5, 2, 3, 5, 3, 2)$ and $opt=2$.

 \section{Conclusion}\label{conclusion}
    In this paper, we study the partial inverse min-max spanning tree problem
    under the weighted sum Hamming distance.
    By considering the alternative set of $w^*(F)$ and restricted version of PInvMMST$_{WSH}$,
    we provide a polynomial time algorithm with time complexity $O(|V|^4)$ to solve this problem.
    Combining this result with previous conclusions,
    the computational complexities of PInvMMST under the classical norms have been fully solved.

\end{document}